\documentclass[useAMS,usenatbib]{mn2e}

\usepackage{natbib, aas_macros}
\usepackage{ulem}
\usepackage{ulem}
\usepackage[dvipdfmx]{graphicx} 
\usepackage{wrapfig}
\graphicspath{{./figs/}}
\usepackage{color}
\usepackage{amsmath}
\usepackage{amssymb}

\newcommand{\Msun}{{\rm  M_{\odot}}}

\newcommand{\Zsun}{Z_{\odot}}

\newcommand{\solmass}{\rm M_{\odot}}

\providecommand{\adsurl}[1]{\href{#1}{ADS}}

\newcommand{\Lya}{$\rm{Ly} \alpha$ }

\newcommand{\HII}{$\rm{H_{II}}$ }

\title[Globular cluster formation under Ly$\alpha$ radiation] 
{Suppression of globular cluster formation in metal-poor gas clouds by Lyman-alpha radiation feedback}
\author[Abe et al. ]{Makito Abe$^{1}$\thanks{E-mail: mabe@ccs.tsukuba.ac.jp} and Hidenobu Yajima$^{2,}$$^{3}$ \\
$^{1}$Center for Computational Sciences, University of Tsukuba, Ten-nodai, 1-1-1 Tsukuba, Ibaraki 305-8577, Japan\\
$^{2}$Frontier Research Institute for Interdisciplinary Sciences, Tohoku University, Sendai 980-8578, Japan\\
$^{3}$Astronomical Institute, Tohoku University, Sendai 980-8578, Japan}

\begin{document}

\date{Accepted ?; Received ??; in original form ???}

\pagerange{\pageref{firstpage}--\pageref{lastpage}} \pubyear{2008}

\maketitle

\label{firstpage}

%
%
\begin{abstract}
We study the impact of \Lya radiation feedback on globular cluster (GC) formation. 
In this Letter, we analytically derive the relation between star formation efficiency (SFE) and metallicity
in spherical clouds with the \Lya radiation feedback. 
Our models show that the SFE becomes small as the metallicity decreases.
In metal-poor gas clouds, \Lya photons are trapped for a long time and exert strong radiation force to the gas,
resulting in the suppression of star formation.
We find that bound star-clusters (${\rm SFE} \gtrsim 0.5$) form only for the metallicity higher than $\sim 10^{-2.5}~\Zsun$
in the case with the initial cloud mass $10^{5}~\Msun$ and the radius $5~\rm pc$.
Our models successfully reproduce the lower bound of observed metallicity of GCs. 
Thus, we suggest that the \Lya radiation feedback can be essential in understanding the formation of GCs. 
\end{abstract}
%
%

%
%
\section{Introduction}

Star formation in the early Universe is poorly known due to little direct observations. 
Globular clusters (GCs) are key objects to understand the early Universe, 
because most of them were likely formed more than $\sim 10$ billion years ago \citep[e.g.,][]{Krauss&Chaboyer03, Dotter+07, VandenBerg+13}.
However, the formation mechanism of GCs has not been understood yet. 
Very recently, \citet{Vanzella+17} observed a young and compact object at $z\sim 6$ of which the physical properties were similar to that of GCs \citep[see also][]{Bouwens+17}. 
In addition, \citet{Renzini17} discussed the detectability of GC progenitors in the early Universe by $\it James$ $\it Webb$ $\it Space$ $\it Telescope$ ($\it JWST$). 
Therefore, the theoretical study of physical conditions for the GC formation is timely just before the JWST era. 
An interesting feature of observed GCs is that the typical metallicity is lower than the solar neighbourhood, but limited at ${\rm log} Z/Z_{\odot} \gtrsim -2.5$ \citep{Puzia+05}.
On the other hand, single metal-poor stars show a wide metallicity range extended to ${\rm log} Z/Z_{\odot} < -4$ \citep{Aoki+07, Yong+13}.
In this work, we study the origin of the lower bound of the observed metallicity of GCs and key physics regulating their formations.


Theoretically, cosmological $N$-body simulations successfully reproduced the distribution of GCs in our Milky Way with sub-grid models \citep{Diemand+05, Moore+06, Trenti+10}. 
High-resolution cosmological hydrodynamics simulations showed that merging process of high-redshift dwarf galaxies induced the formation of compact clouds where were likely to be formation sites of GCs
\citep[e.g.,][]{Kravtsov&Gnedin05,Ricotti+16,Kim+17}. 
\citet{Abe+16} suggested that compact gas clouds were also formed under the strong ultraviolet (UV) background radiation
by using the radiative-hydrodynamics simulations \citep[see also,][]{Hasegawa2009}.
Thus, recent hydrodynamics simulations have been able to resolve the formation of compact gas clouds. 
On the other hand, the formation mechanism of GCs in the compact clouds remains unknown. 
Once stars form, the internal stellar feedback can disrupt the clouds, resulting in suppression of star formation \citep[e.g.,][]{Geen+17}.
Therefore, the internal feedback has to be considered to understand the formation of GCs.

We here consider \Lya radiation as the internal feedback.  
Due to its large cross-sections and the resonant scattering nature, \Lya photons would be trapped for a long time in a star-forming cloud and exert strong radiation force to the gas \citep{Rees&Ostriker77, Cox85, Oh&Haiman02, Yajima&Khochfar14, Yajima&Khochfar17}.
Such \Lya radiation feedback in the high-$z$ star-forming galaxies has been argued by using radiative transfer simulations \citep{Dijkstra&Loeb08}, or one-dimensional \Lya radiative-hydrodynamics simulations \citep{Smith+17}. 
These works revealed that the \Lya radiation feedback induced the outflow of gas from the low-mass protogalaxies with the halo masses of $M_{\rm h}\lesssim 10^8~\solmass$. 
This can cause the quench of star formation. 
Therefore, \Lya radiation feedback may also play a roll even in the formation of GCs.

On the other hand, as the gas is metal-enriched, \Lya photons are absorbed by dust in the cloud. 
Even in low-metal gas clouds, the dust reduces the number density of trapped \Lya photons, 
and make the \Lya feedback ineffective. 
The impacts of \Lya radiation feedback on the GC formation in low-metallicity gas clouds has not been investigated. 
In this work, we study the conditions for the formation of GCs by taking into account the \Lya radiation feedback with the dust absorption. 




%
%
\section{Star formation in spherical gas clouds under the \Lya radiative feedback}
In this work, we consider star formation in spherical gas clouds with the mass $M_{\rm cl}$ and the radius $r_{\rm cl}$.
As the star formation proceeds, massive stars form \HII bubbles in a cloud. 
In the \HII region, \Lya photons are produced via recombination processes, and can exert radiation force on the cloud. 
We study the regulation of star formation due to the \Lya radiation force under the assumption of a single point source at the centre of the cloud, and derive the physical conditions for the GC formation. 

In the equilibrium state, \Lya luminosity is simply proportional to ionizing photon emissivity as
$		L_{_{{\rm Ly} \alpha}} =  0.68 \epsilon_{\rm Ly\alpha} \dot{N}_{\rm ion}$, 
where $\epsilon_{\rm Ly\alpha}=10.2 ~\rm eV$ is the energy of each \Lya photon, and $\dot{N}_{\rm ion}$
is the ionizing photon emissivity from a star cluster \citep{ART2}. 
When the age of star cluster is younger than the lifetime of massive stars, 
the ionizing photon emissivity is proportional to the stellar mass $M_\ast = \epsilon M_{\rm cl}$, where $\epsilon$ denotes the star formation efficiency (SFE) of the cloud.
Under the assumption of the Salpeter initial mass function (IMF) with the mass range of $0.1 - 100~\solmass$ and the metallicity of $Z/Z_{\odot} = 0.01$, the ionizing photon emissivity
per solar mass $\dot{n}_{\rm ion}$ is evaluated as $\sim 4.1\times 10^{46}~{\rm s}^{-1} \solmass^{-1}$ at the age of $1~\rm Myr$ \citep{Chen+15}. 
On the other hand, if we adopt the Larson IMF with its characteristic mass $M_{\rm ch} =1~\solmass$, the ionizing photon emissivity somewhat increases, resulting in $\dot{n}_{\rm ion} \sim 1.1\times 10^{47}~{\rm s}^{-1} \solmass^{-1}$. 
Thus, we evaluate the \Lya luminosity in the cloud as 
\begin{eqnarray}
	\label{eq:L_Lya}
		L_{_{{\rm Ly} \alpha}} = 
		1.1 \times 10^{36} ~{\rm erg~s^{-1}}  ~\epsilon \left( \frac{\dot{n}_{\rm ion}}{1\times 10^{47}~\rm s^{-1}\solmass^{-1}} \right) \left(\frac{M_{\rm cl}}{\solmass}\right). 
\end{eqnarray}

Next, we estimate the \Lya radiation force enhanced due to multiple scattering processes as \citep{Dijkstra&Loeb08} 
\begin{equation}
	\label{eq:F_Lya}
		F_{\rm rad}^{\rm Ly \alpha}\sim \frac{t_{\rm trap}}{t_{\rm cross}}\frac{L_{{\rm Ly} \alpha}}{c} \equiv f_{\rm boost} \frac{L_{{\rm Ly} \alpha}}{c},  
\end{equation}
where $c$ is the speed of light, $t_{\rm trap}$ is the trapping time of \Lya photons in a cloud, $t_{\rm cross} = r_{\rm cl} / c$ is the light crossing time, and 
\begin{equation}
	\label{eq:fboost1}
	f_{\rm boost} = \frac{t_{\rm trap}}{t_{\rm cross}} = \frac{c}{r_{\rm cl}}t_{\rm trap}
\end{equation}
is the boost factor of the radiation force due to multiple scattering processes. 
According to the numerical study \citep{Adams75},  the trapping time {$t_{\rm trap, H}$} can be expressed by the optical depth at the line center frequency $\tau_0$ as 
\begin{equation}
\label{eq:fboost}
t_{\rm trap,H} \sim \begin{cases}
 15 \; t_{\rm cross} ~~&{\rm for}~10^{3} < \tau_{0} < 10^{5.5}\\
15 \; \left( \frac{\tau}{10^{5.5}} \right)^{\frac{1}{3}} t_{\rm cross}  ~~&{\rm for}~\tau_{0} \ge 10^{5.5}.
\end{cases}
\end{equation}

As the photon density increases, the \Lya radiation force pushes the cloud outward and prevents it to contract. 
When the \Lya radiation force exceeds the gravitational force, i.e., $F_{\rm Ly\alpha} > F_{\rm grav}$, 
the gas is likely to be evacuated, resulting in the quenching of star formation. 
This condition is described as
\begin{equation}
	\label{eq:condition}
		 f_{\rm boost} \frac{L_{{\rm Ly} \alpha}}{c} > G\frac{(1-\epsilon) M_{\rm cl}^2}{r_{\rm cl}^2}. 
\end{equation}
Thus, by substituting Eq. (\ref{eq:L_Lya}) into Eq. (\ref{eq:condition}), we obtain the boost factor to suppress the star formation as a function of $\xi \equiv \epsilon/ (1-\epsilon) $ (corresponding to ``stellar-to-gas mass ratio'') :  
\begin{equation}
	\label{eq:fboost_condition}
		f_{\rm boost} > 3.0~ \xi^{-1} \left( \frac{\dot{n}_{\rm ion}}{1\times 10^{47}~\rm s^{-1}\solmass^{-1}} \right)^{-1} 
		\left(\frac{M_{\rm cl}}{10^5 ~\solmass}\right)\left(\frac{r_{\rm cl}}{5~{\rm pc}}\right)^{-2}.
\end{equation}

In order to estimate $f_{\rm boost}$, we calculate the trapping time of the \Lya photons. 
The optical depth from the cloud centre to the edge is estimated as
\begin{equation}
	\label{eq:NH}
		\tau_{0} = 7.1\times 10^{9}~\left(\frac{M_{\rm cl}}{10^5~\solmass}\right)\left(\frac{r_{\rm cl}}{5~{\rm pc}}\right)^{-2}.
\end{equation}
By considering the condition $\tau_{0} > 10^{5.5}$ in Eq. (\ref{eq:fboost}), this results in
\begin{eqnarray}
	\label{eq:t_trap}
	     t_{\rm trap, H} =6.9\times10^3~{\rm yr}  \left(\frac{M_{\rm cl}}{10^5~\solmass}\right)^{\!\!1/3}\!\! \left(\frac{r_{\rm cl}}{5~{\rm pc}}\right)^{\!\!1/3}\!\!. 
\end{eqnarray}
Thus, according to Eq. (\ref{eq:fboost1}) and (\ref{eq:t_trap}), the boost factor is given by
\begin{eqnarray}
	\label{eq:fboost_lya}
	     f_{\rm boost,H} =4.2\times10^2  \left(\frac{M_{\rm cl}}{10^5~\solmass}\right)^{\!\!1/3}\!\! \left(\frac{r_{\rm cl}}{5~{\rm pc}}\right)^{\!\!-2/3}\!\!. 
\end{eqnarray}

As the metallicity increases, the trapping time can be shorter than the estimation in Eq.~(\ref{eq:fboost})
by taking into account the dust absorption of \Lya photons during the journey.
The mean free path of the \Lya photons regarding the dust absorption $l_{\rm d}$ is $l_{\rm d} = 1/(\sigma_{\rm d} n_{\rm d})$, where 
\begin{equation}
	\label{eq:tau_d}
	\sigma_{\rm d} n_{\rm d} =\pi a_{\rm d}^2 Q_{\rm d, abs} n_{\rm H} \left(\frac{m_{\rm H}}{m_{\rm d}}\right) f_{\rm d} \left(\frac{Z}{Z_\odot}\right). 
\end{equation}
Here, $a_{\rm d}$ represents the dust size, $Q_{\rm d, abs}$ is the absorption efficiency to the geometrical cross section, $f_{\rm d}$ is the dust-to-gas mass ratio at the metallicity of solar abundance, $n_{\rm H}$ and $m_{\rm H}$ are hydrogen number density and hydrogen mass. 
We set $Q_{\rm d, abs}=1$, which is reasonable when the wavelength of photons is shorter than $\sim 2 \pi a_{\rm d}$ \citep{Draine&Lee84}. 
As a fiducial value, we set $f_{\rm d} = 0.01$ that is similar to the value at solar neighborhood \citep{Spitzer78}. 
In this work, we assume that the dust-to-gas mass ratio is proportional to the metallicity \citep{Draine+07}.
The dust mass $m_{\rm d}$ is given by $m_{\rm d} = 4\pi a_{\rm d}^3 \rho_{\rm d}/3$, where $\rho_{\rm d}$ is the dust mass density. 
We assume $\rho_{\rm d} = 3~\rm g ~ cm^{-3}$ that is corresponding to a silicate dust grain. 
The dust size distribution in high-redshift galaxies has not been understood yet. 
In our Galaxy, the dust is likely to have a power-law size distribution $\frac{dn_{\rm d}}{da_{\rm d}} \propto a_{\rm d}^{-3.5}$ with the size range $\sim 0.01 - 1.0~\rm \mu m$, so called the MRN size distribution \citep{Mathis+77}. 
In this work, we adopt a single dust size model with a fiducial value $a_{\rm d}=0.1~\rm \mu m$. 
Note that, this single dust-size model is equivalent with the MRN dust model with the size range $8\times10^{-3} - 1.0~\rm \mu m$ under the condition $Q_{\rm d, abs}=1$ \citep{Yajima+17}. 

\begin{figure}
	\begin{center}
		\includegraphics[width=9.5 cm,clip]{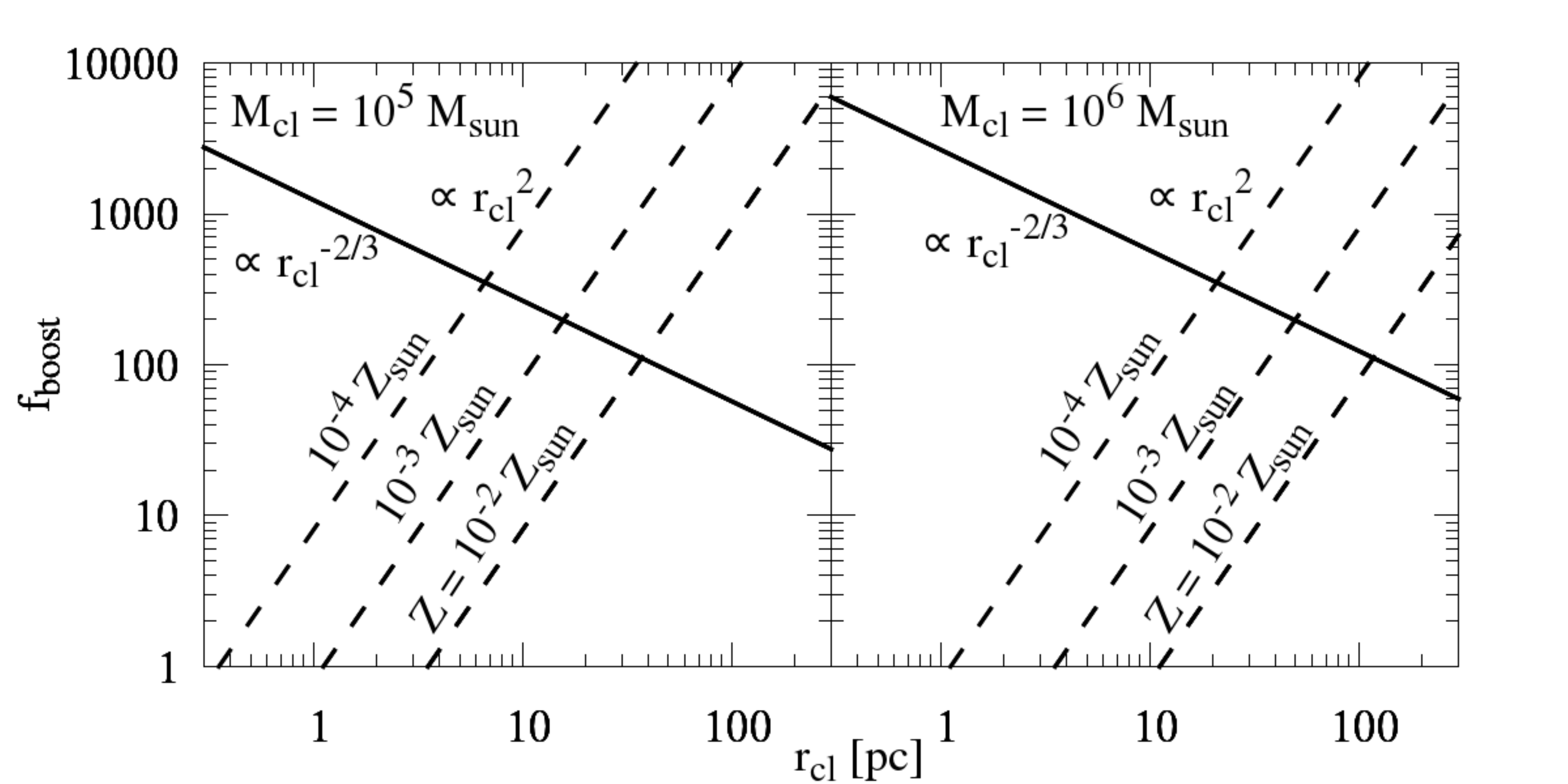}
	\end{center}
	\vspace{-5mm}
	\caption{
	The boost factor as a function of cloud radius. 	 
	Solid lines show the cases in which the dust absorption is not considered. 
	Dash lines represent boost factors calculated by the trapping time 
	until the optical depth of dust absorption becomes unity. 	
	}  
	\label{fig:fboost}
\end{figure}

Thus, the traveling time $t_{\rm d} = l_{\rm d}/c$ is given by
\begin{eqnarray}
	\label{eq:t_dust}
		t_{\rm d} \!\!\!\! &\sim& \!\!\!\!
		33~{\rm yr} \left( \frac{M_{\rm cl}}{10^5~\solmass} \right)^{\!\!\!-1}\!\!\! \left(\frac{r_{\rm cl}}{5~{\rm pc}}\right)^{3}\!\!\!
		\left( \frac{Z}{10^{-2}~Z_\odot} \right)^{\!\!-1}\!\!\! \left(\frac{X_{\rm d}}{0.1~\mu {\rm m}^{-1}} \right)^{-1}\!\! , \nonumber \\
\end{eqnarray}
here we define $X_{\rm d}  \equiv f_{\rm d} / a_{\rm d}$. 
When we consider the photon trapping time as $t_{\rm d}$ and substitute Eq. (\ref{eq:t_dust}) into Eq. (\ref{eq:fboost1}), the boost factor is evaluated as 
\begin{eqnarray}
	\label{eq:value_fboost}
	f_{\rm boost,d} 
	 =  2.0 \left(\frac{M_{\rm cl}}{10^5~\solmass}\right)^{\!\!-1}\!\!\!  \left(\frac{r_{\rm cl}}{5~{\rm pc}}\right)^{\!\!2}\!\!\! 
	 \left(\frac{X_{\rm d}}{0.1~\rm {\mu m^{-1}}}\right)^{\!\!-1}\!\!\! \left(\frac{Z}{10^{-2}~Z_\odot}\right)^{\!\!-1}.  \!\!\!\!\!\!\!\! \nonumber \\ 
\end{eqnarray}
We estimate the trapping time of \Lya photons as $t_{\rm trap} = {\rm min}(t_{\rm d},~t_{\rm trap, H})$. 
The critical metallicity that satisfies $t_{\rm trap, H} = t_{\rm d}$ is 
\begin{equation}
	\label{eq:Z_crit}
	Z_{\rm TP} = 4.7\times 10^{-5} Z_\odot \left(\frac{M_{\rm cl}}{10^5~\solmass}\right)^{\!\!-4/3}\!\! \left(\frac{r_{\rm cl}}{5~{\rm pc}}\right)^{\!\!8/3} \!\!\! \left(\frac{X_{\rm d}}{0.1~\mu {\rm m}^{-1}} \right)^{-1}.
\end{equation}
Note that, the momentum of \Lya photons is given to H{\sc i} gas or dust at least once. 
Therefore, we take a higher value of the above estimation or unity as a minimum of $f_{\rm boost}$. 
Fig.~\ref{fig:fboost} shows the boost factor as a function of cloud radius. 
The dependence of cloud radius on the boost factor changes from $\propto r^{2}$ to $\propto r^{-2/3}$ at a specific radius 
that increases with metallicity as shown in Eq. (\ref{eq:Z_crit}).
We find that the boost factor is regulated by the dust in the cases of compact clouds of which the mass and radius are similar to GCs. 
Therefore, we consider the boost factor as $f_{\rm boost} = f_{\rm boost,d}$ in cases with the metallicity higher than $Z_{\rm TP}$. 

Substituting Eq. (\ref{eq:value_fboost}) into Eq. (\ref{eq:fboost_condition}),
we finally obtain the critical value for the stellar-to-gas mass ratio ($\xi_{\rm crit}$) above which the \Lya radiation feedback suppresses the star formation:
\begin{equation}
\begin{split}
	\label{eq:xi-Z}
	\xi_{\rm crit} = 1.5
	\left( \frac{\dot{n}_{\rm ion}}{1\times 10^{47}~\rm s^{-1}~\solmass^{-1}} \right)^{-1}  \!
	\left(\frac{M_{\rm cl}}{10^5~\solmass}\right)^{\!\!2}\!  \left(\frac{r_{\rm cl}}{5~{\rm pc}}\right)^{\!\!-4}\!\!\!  \\
	                    \times  \left(\frac{X_{\rm d}}{0.1~\rm {\mu m^{-1}}}\right)\!\ \left(\frac{Z}{10^{-2}~Z_\odot}\right). 
\end{split}
\end{equation}
The above estimation suggests that the star formation efficiency is regulated due to the \Lya radiative feedback 
as the metallicity decreases. Below, we investigate the critical condition for the metallicity.  

If most of gas is evacuated due to the feedback, 
stars are likely to escape from the shallowed potential well. 
\citet{Hills80} analytically showed that star clusters became unbound if $\epsilon \lesssim 0.5$  \citep[see also,][]{Mathieu83}.
Then, \citet{Geyer&Burkert01} confirmed this critical condition by $N$-body simulations. 
Therefore, by considering the critical star formation efficiency as $\epsilon = 0.5$ (i.e., $\xi = 1$), 
we evaluate the critical metallicity $Z_{\rm crit}$ for forming bound star-clusters from Eq. (\ref{eq:xi-Z}):
\begin{equation}
\begin{split}
	\label{eq:bound_Z}
	Z_{\rm crit} = 6.6 \times10^{-3} Z_\odot \left( \frac{\dot{n}_{\rm ion}}{1\times 10^{47}~\rm s^{-1}~\solmass^{-1}} \right)
	 \left(\frac{X_{\rm d}}{0.1\rm {\mu m^{-1}}}\right)^{-1} \\
	 \times \left(\frac{M_{\rm cl}}{10^5~\solmass}\right)^{\!\!-2}\!\!\!  \left(\frac{r_{\rm cl}}{5~{\rm pc}}\right)^{\!\!4} . 
\end{split}
\end{equation} 

Note that there are uncertainties in modeling the parameter $X_{\rm d}$. 
\citet{Nozawa+07} showed that the size distribution of dust in the early Universe tended to be biased toward the large size ($\gtrsim 0.1~{\rm \mu m}$) since small grains were destroyed by the reverse shock of the supernova (SN) explosion. 
If we consider the dust size range as $0.1~{\rm \mu m} < a_{\rm d} < 1.0 ~{\rm \mu m}$, the corresponding single dust size is $\sim 0.3~{\rm \mu m}$. 
On the other hand, the typical dust size of the high-$z$ galaxies has also been argued as smaller as $0.05~{\rm \mu m}$ \citep{Todini&Ferrara01, Dayal+10}. 
Thus, $X_{\rm d}$ can be changed by a factor $\sim 0.5 - 3$. 

In addition,  the dust-to-gas mass ratio $f_{\rm d}$ can also change since the dust depletion factor in the early Universe (the mass ratio of dust to the total heavy elements) may be different compared to the local Universe. 
\citet{Schneider+12} estimated that the depletion factor as $\sim 3.4 \times 10^{-2}$ {\rm for the SN with the progenitor mass of 35~$\solmass$ and the metallicity of $Z/Z_\odot = 10^{-4}$}, which was roughly corresponding to the tenth of the present-day depletion factor \citep{Pollack+94, Schneider+12}. 
Consequently, our dust parameter $X_{\rm d}$ can be $\sim$3 times larger and/or $\sim$10 times smaller than $X_{\rm d} = 0.1~{\rm \mu m}^{-1}$. 

\begin{figure}
	\begin{center}
		\includegraphics[width=8.5 cm,clip]{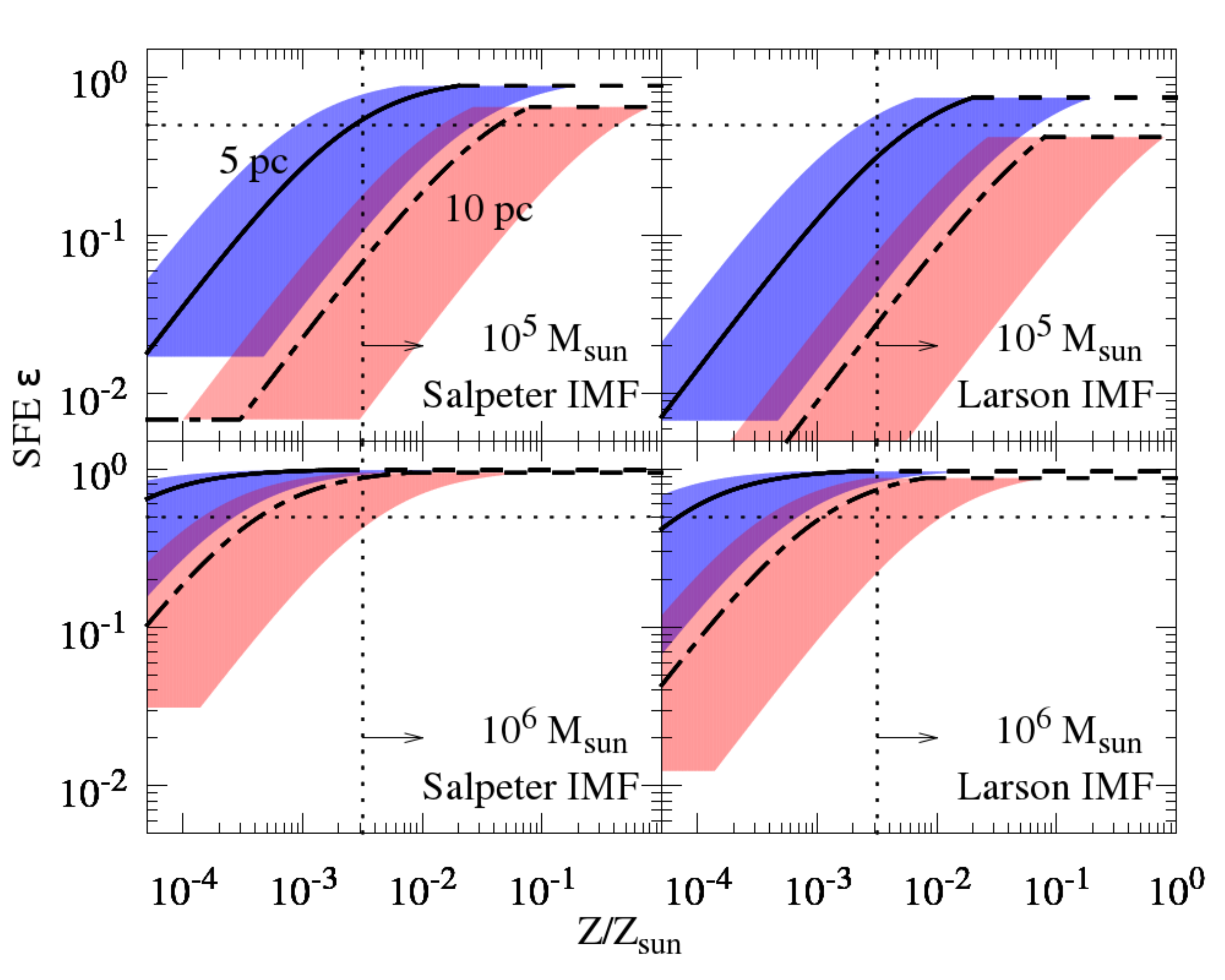}
	\end{center}
	\vspace{-5mm}
	\caption{Critical star formation efficiency $\epsilon$ as a function of the metallicity $Z$, obtained by the Eq. (\ref{eq:xi-Z}). 
		The star formation is suppressed by \Lya radiation feedback if the SFE exceeds the line.
		In each panel, solid and dashed-dotted lines correspond to the cloud radius of $r_{\rm cl} = 5~{\rm pc}$ and $10~{\rm pc}$, respectively. 
		As for each line, the fiducial dust model of $X_{\rm d} = 0.1~{\rm \mu m^{-1}}$ is adopted. 
		The shaded regions represent the uncertainty of the dust model (see the text in detail). 
		Dashed lines represent the minimum boost factor of $f_{\rm boost} = 1$. 
		Upper panels show the results for $M_{\rm cl} = 10^5~\solmass$ and lower panels are $M_{\rm cl} = 10^6~\solmass$. 
		Left-hand and right-hand side panels show the results for the Salpeter and Larson IMF, respectively. 
		Vertical dotted lines denote the lower bound of the observed metallicity of GCs, $\log Z/Z_\odot = -2.5$. 
		The horizontal dotted lines represent $\epsilon = 0.5$. 
		Note that the SFR becomes constant at $Z < Z_{\rm TP}$ because of the constant traveling time of \Lya photons. 
		  }
	\label{fig:epsilon-Z}
\end{figure}
Fig. \ref{fig:epsilon-Z} shows the critical 
star formation efficiency as a function of metallicity. 
Shades represent the uncertainty of $X_{\rm d}$ as stated above, i.e., $0.01 < X_{\rm d} < 0.3$. 
The dashed line parts represent $f_{\rm boost}=1$. 
In the case with the mass $10^5~\solmass$ and the radius 5~pc, the SFE becomes smaller than $\sim 0.5$ for Salpeter IMF at the metallicity of $\log~Z/Z_\odot \sim -2.5$, 
which nicely reproduces the lower bound of observed metallicity of GCs. 
Note that, for Larson IMF, the {\rm critical} metallicity shifts higher side by a factor $\sim 2$ due to the higher production rate of \Lya photons. 
As the metallicity becomes lower than $\log Z/Z_{\odot} \sim -2.5$, 
\Lya photons are able to travel for a long time with multiple scatterings and exert the outward force on the gas, resulting in the small SFE. 
For the metallicity of $\log Z/Z_{\odot} = -4$ which is similar to observed metal-poor stars, the SFE becomes $\epsilon \sim 3.6 \times 10^{-2}$ due to the large boost factor  $f_{\rm boost} \sim 200$. 
Therefore, stars formed in metal-poor gas are likely to disperse. 
Thus we suggest that GCs are difficult to form in metal-poor gas clouds. 
As the size of the cloud increases, i.e., {\rm the} gas density decreases, the critical SFE becomes smaller because of the longer traveling time (see Eq. \ref{eq:t_dust}) and the shallower gravitational potential. 

When the boost factor reaches unity, the critical SFE becomes constant to the metallicity as shown in Eq. (\ref{eq:fboost_lya}). 
In the case with the cloud mass $10^6~\solmass$, 
\Lya photons are absorbed by dust before they exert the strong radiation force on the gas, 
and the deeper gravitation potential can hold the gas against the feedback.
As a result, 
the \Lya radiation feedback is insignificant
even in the low-metallicity regions while the \Lya feedback begins to work (i.e., $f_{\rm boost} > 1$) at the metallicity of  $\log Z/Z_\odot \sim -2.5$. 
If we consider the radius of 10~pc and Larson IMF, 
the result seems to be reasonable to explain the observed lower bound of metallicity. 

\begin{figure}
	\begin{center}
		\includegraphics[width=8.5cm,clip]{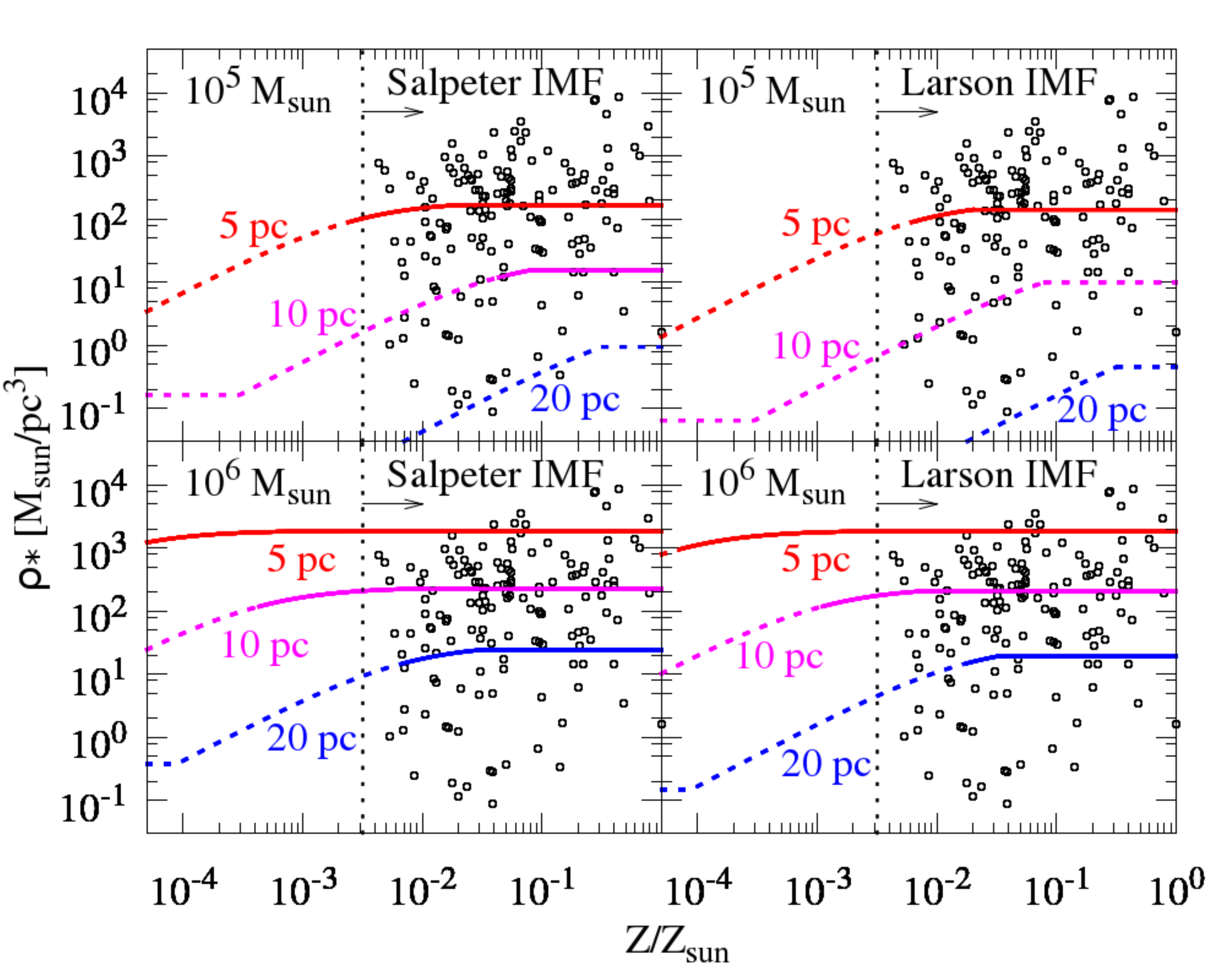}
	\end{center}
	\vspace{-5mm}
	\caption{Stellar densities of star clusters as a function of the metallicity $Z$ for each model of Fig. \ref{fig:epsilon-Z}. 
		Same as Fig. {\ref{fig:epsilon-Z}}, vertical dotted lines denote the metallicity of $\log Z/Z_\odot = -2.5$. 
		In each panel, solid, dashed-dotted and dashed-two dotted lines represent results for clouds with $r_{\rm cl} = 5~{\rm pc}$, $10~{\rm pc}$ and $20~{\rm pc}$. 
		Open circles indicate the constant half-mass density $\rho_{\rm h} \equiv 3M_{\ast}/8\pi r_{\rm h}^3$ \citep{PortegiesZwart+10} for observed GCs \citep{Harris1996}. 
		In this figure, we assume the fiducial dust model of $X_{\rm d} = 0.1~{\rm \mu m^{-1}}$. 		
		 }
	\label{fig:rho-Z}
\end{figure}

Next, we estimate the stellar density as 
$\rho_\ast \sim \frac{3M_\ast}{4\pi r_{\rm cl}^3} = \frac{3\epsilon M_{\rm cl}}{4\pi r_{\rm cl}^3}. $
By considering the metallicity dependence of the star formation efficiency (Eq. \ref{eq:xi-Z}), 
we derive the stellar mass density as a function of metallicity as shown in Fig. \ref{fig:rho-Z}.
As the reference, we present the constant half-mass stellar density $\rho_{\rm h} \equiv 3M_{\ast}/8\pi r_{\rm h}^3$ for observed GCs \citep{PortegiesZwart+10}, assuming a mass-to-light ratio $M_\ast / L_{\rm V} = 2$ \citep{Pryor&Meylan93} and ${\rm [Fe/H]} = Z/Z_\odot$. 
The observational data are taken from \citet{Harris1996}. 
Dotted lines represent the cases with $\epsilon < 0.5$. 
In these cases, stars become unbound after the gas evacuation. 
Therefore, only solid line parts should be considered in studying the stellar mass density.
Note that, the estimated stellar density 
is the lower limit since the star clusters generally have the stellar density profile. 
For instance, if we assume the isothermal density profile ($\rho(r) \propto r^{-2}$ hence $M_\ast(r)\propto r$), the stellar density becomes four times larger if we estimate the stellar density at the half-mass radius.  
We see in the panels that the stellar densities decrease with the metallicity. 
This feature arises from the decreasing critical SFE, and the masses of formed star clusters become lower at given radii.
We show some cloud models successfully reproduce the observed stellar density of GCs at the metallicity higher than $\log Z/Z_\odot \sim -2.5$, and the lower bound of the metallicity. 

\section{Discussion and Conclusions}

In this Letter, we study the star formation in compact gas clouds under \Lya radiation feedback.
We find that the \Lya radiation feedback significantly suppresses the star formation when the metallicity of the cloud is low. As the metallicity decreases, \Lya photons can be trapped in the cloud for a long time against dust absorption.
In the journey of \Lya photons, they can exert the strong radiation pressure on the gas via multiple scattering processes, resulting in the disruption of clouds.
If the star formation efficiency is smaller than $\sim 0.5$, stars become unbound after the gas evacuation due to the radiation pressure or supernovae. 
Therefore, star formation is suppressed in metal-poor gas clouds, 
and compactly bound star clusters are unlikely to form. 
In the case of a cloud with the mass $10^{5}~M_{\odot}$ and the radius $5~\rm pc$, 
we show the critical metallicity for forming a bound star cluster is ${\rm log}\; Z/Z_{\odot} \sim - 2.5$
above which more than half of gas can convert into stars.
This critical metallicity is similar to the lower bound of the observed metallicity of globular clusters (GCs).
Thus we suggest that the \Lya radiation feedback can be a key roll in the formation of GCs.

 In this work, we assume the simple dust model with the typical size of $0.1~\rm \mu m$
 and the constant dust-to-metal mass ratio normalized by that of the solar neighborhood.  
 However, in the early Universe, the dust properties can significantly differ from local galaxies. 
 This dust properties can also affect the formation of cloud properties
 due to the different thermal evolution of low-metal gas \citep[e.g.,][]{Omukai+05}. 
 We will investigate the formation of dusty gas clumps by using cosmological hydrodynamics simulations with 
 the evolution of dust properties in future work. 
Our models are based on spherical gas clouds that are likely to trap \Lya photons efficiently. 
If Lyman continuum radiation from a star cluster makes ionized holes along low-density regions, \Lya photons can leak and the radiation force becomes weak. 
In addition, supersonic motion of gas, which is by turbulence or outflow, 
make the trapping time of \Lya photons shorter. 
These should be studied by multi-dimensional radiative-hydrodynamics simulations with 
Lyman continuum and \Lya line transfer calculations.


%
%
\section*{Acknowledgments}
We are grateful to H. Fukushima for fruitful discussion about the ionizing photon emissivity. 
We also thank the anonymous referee for useful comments that have improved the manuscript. 
This work is supported by MEXT/JSPS KAKENHI Grant Number 17H04827 (H.Y.). 

%
%
\bibliographystyle{mn}
\bibliography{mn-jour,ref-bibtex}

\label{lastpage}

\end{document}